# Leveraging learning analytics to enhance immersive teacher simulations: Challenges and opportunities[1]

Sumin Hong, Seoul National University, Seoul 08826, Republic of Korea, hongxiumin@snu.ac.kr
Jewoong Moon, The University of Alabama, AL 35401, USA, jmoon19@ua.edu
Taeyeon Eom, Seoul National University, Seoul 08826, Republic of Korea, eomcalm@gmail.com
Juno Hwang, Seoul National University, Seoul 08826, Republic of Korea, wnsdh10@snu.ac.kr
Jibeom Seo, Seoul National University, Seoul 08826, Republic of Korea, jabam1264@snu.ac.kr

**Abstract**

This chapter examines how data analytics can be leveraged to enhance immersive teacher simulations, situating this inquiry within the broader learning sciences discourse on embodied cognition, data-informed feedback, and teacher professional learning. It explores both conceptual foundations and empirical cases to illustrate how analytics serve as mediational tools that connect immersive experiences with reflective teaching practice. The chapter unfolds in multiple sections: (1) *The Innovation Journey: An Overview of Immersive Teacher Simulations* outlines the evolution from traditional simulations to XR-based environments, highlighting the need for professional decision-making under realistic constraints. (2) *Innovation in Existing Research and Practice* situates teacher analytics within the trajectory from descriptive observation to multimodal and predictive modeling. (3) *Study Approach and Design* details how multimodal data—discourse, behavior, and gaze—from the TeacherGen@i simulation were collected and organized to reveal cognitive distribution of pedagogical discourse and interaction patterns. (4) *Findings* present the cognitive distribution of preservice teachers' pedagogical discourse and the sequential interaction patterns that emerge in exchange, illustrating how multimodal analytics make pedagogical reasoning processes visible within immersive simulations. (5) *Understanding Innovative Practices in Teacher Education* examines teaching analytics to enhance immersive teacher simulation based on the findings of the study. (6) *Key Takeaways of the Innovation Journey* identifies research challenges and design implications for scalable, analytics-enhanced teacher education. Together, these sections position immersive teacher simulations as a pivotal

---

testbed for aligning learning analytics, professional learning, and next-generation immersive learning environment design.

**The Innovation Journey: An Overview of Immersive Teacher Simulations**

In teacher education, simulation-based learning has been utilized as an important means to help both preservice and in-service teachers rehearse decision-making, reflection, and adaptive expertise in controlled yet authentic contexts (Kaufman & Ireland, 2016; Ledger et al., 2025). Traditional methods, such as case-based learning, role-play, and microteaching, have contributed to the development of lesson planning and problem-solving skills; however, they often lack sensorimotor engagement and ecological validity, which are key features that underlie authentic teaching experiences. In particular, traditional simulations are often limited to a small number of students, making it difficult to reproduce the diverse reactions and situations that occur in classrooms, as well as the complexity and unpredictability of real teaching. This, in turn, can constrain teachers' ability to apply knowledge and skills effectively in authentic educational environments.

Immersive teacher simulations address these gaps by situating practice within high-fidelity virtual or mixed-reality environments that replicate classroom contingencies with greater ecological realism. There has been a transition from simple video-based simulations to more sophisticated mixed-reality simulations such as *TeachLivE™* and *simSchool* (Dieker et al., 2023). They enable teachers to engage in repeated trials of the same scenario, test alternative responses, and observe system-level consequences of their actions, which collectively enhance the precision of professional judgment. In this sense, immersion is not merely a matter of sensory realism but a design principle that supports cognitive realism, indicating the alignment between simulated and real-world design processes (Badiee & Kaufman, 2015; Codreanu, E. et al., 2020).

This integration of emerging technologies, such as VR and AR, and generative artificial intelligence (GenAI) has further extended the reach of simulation-based teacher education. GenAI agents now allow dynamic, unscripted interactions that approximate the fluid discourse patterns of real students, creating richer opportunities for responsive teaching and situational awareness (Docter et al., 2024; Hong et al., 2025; Hwang et al., 2024). These developments collectively signal a methodological and conceptual evolution in how simulations support teacher learning. Immersive environments enhanced with AI capabilities now operate as adaptive ecosystems that capture, interpret, and respond to teachers' decisions in real time. This responsiveness provides not only heightened authenticity but also generates a continuous stream of behavioral and linguistic data. In this way, immersive simulations have begun to converge with the growing field of teacher analytics, where data-informed insights are used to understand and enhance professional decision-making. The capacity to record and analyze teachers' micro-level actions—such as question patterns, feedback timing, gaze shifts, or tone modulation—has expanded the analytic scope of teacher education (Marelle et al., 2025). By embedding such analytic capabilities within immersive environments, simulations can move beyond static role-play toward evidence-informed cycles of reflection and improvement

However, as Dieker et al. (2023) caution, these advancements also underscore an ongoing challenge: while simulations have evolved in technical sophistication, systematic frameworks for evaluating teaching behaviors within them remain limited. Addressing this gap requires integrating analytics as a core design dimension of immersive teacher simulations, enabling both researchers and practitioners to trace how teachers perceive, decide, and act in dynamic classroom contexts.

Building on this need, the present study uses the generative AI-enhanced immersive teacher simulation TeacherGen@i as a case to propose a conceptual and methodological foundation for analyzing teachers' discourse, behaviors, and visual attention in a multilayered manner. Specifically, we examine how a multimodal analytics framework—integrating natural language processing (NLP), behavioral detection, and gaze tracking—can reveal underlying structures of preservice teachers' pedagogical reasoning and interaction patterns.

Accordingly, this study aims to achieve two primary objectives: first, to develop and demonstrate procedures and analytic tools that systematically capture preservice teachers' pedagogical discourse and behaviors within an immersive simulation; and second, to explore how these multimodal indicators illuminate teachers' pedagogical reasoning, temporal interaction patterns, and visual attention distributions.  To achieve these aims, the study addresses the following research questions:

RQ1. What are the characteristics of preservice teachers' pedagogical discourse in TeacherGen@i, and how do these utterances distribute across Bloom's cognitive taxonomy during simulation-based interactions with AI-driven student agents?

RQ2. What sequential interaction patterns emerge between preservice teachers and AI students, and how do these behavioral flows reflect the structure of pedagogical reasoning?

RQ3. How can a teacher simulation dashboard be designed to visualize multimodal data in ways that support analytic insight and reflective practice?

In the following section, we provide an overview of the conceptual and methodological developments that make this integration both feasible and necessary.

**Innovation in Existing Research and Practice**
*Learning Analytics in Teacher Simulation: A Paradigm Shift*
Teacher analytics has gradually evolved over the past several decades in accordance with advances in evaluation methods and technology. Teacher analytics can be defined as the systematic collection and analysis of data on instructional practices, teacher behaviors, and teaching outcomes to improve instructional quality and professional expertise (Hoyos & Velásquez, 2020; Saar et al., 2022). In its early stages, traditional evaluation methods such as classroom observations, student surveys, and peer supervision were the primary means of assessing teachers' instructional competence. However, these approaches were often limited by a high degree of subjectivity and tended to provide outcome-centered, restricted information rather than capturing the fine-grained processes of teaching practice.

Over time, the analytical focus in teacher education has shifted from summative appraisal to formative and process-oriented insight, mirroring broader transformations in learning analytic research. With the widespread adoption of Learning Management Systems (LMSs), digital tool–based analytics, such as clickstream analysis and online participation tracking, were introduced, enabling more systematic and quantitative collection and analysis of teachers' instructional practices and students' responses (Wise & Jung, 2019). This transition represents a foundational step toward data-informed professional learning, allowing teachers to be trained as a dynamic process. It opened up new possibilities for recording and utilizing educational data in real time and laid the foundation for data-driven decision-making in teacher education and instructional improvement.

Recent advances have further accelerated this evolution, giving rise to multimodal analytics that capture the cognitive, behavioral, and affective dimensions of teaching. Representative examples include gaze tracking, speech analysis, physiological signals, and AI-driven pattern recognition (Hur & Bosch, 2024). Although current applications remain relatively limited, these analytical methods are beginning to illuminate previously hidden aspects of teacher cognition—such as situational awareness, response timing, and attentional focus—thereby contributing to a more nuanced understanding of teaching as an adaptive, data-rich practice. The recent advancement of teacher analytics also reflects a broader shift from descriptive to predictive analytics, moving beyond simply documenting what has happened toward forecasting what might happen. This transition enables teacher educators and practitioners to anticipate instructional challenges, identify at-risk patterns, and intervene proactively. Moreover, the field has witnessed a movement from descriptive to predictive analytics, shifting emphasis from documenting what has occurred to anticipating what might occur in complex instructional scenarios. Predictive modeling enables teacher educators to identify patterns of struggle or success early on, offering new ways to support adaptive expertise and reflective growth. The potential of data-informed prediction is particularly compelling in immersive contexts, where temporal data such as gaze trajectories or verbal response patterns can be analyzed to forecast performance trends or cognitive load indicators in real time.

Another critical advancement concerns the emergence of personalized and data-informed feedback for teachers. By systematically analyzing instructional data, analytics can identify teachers' strengths, diagnose areas for improvement, and generate individualized recommendations for practice. Such data-driven feedback goes beyond generic assessment to foster reflective inquiry and evidence-based professional development. When embedded within immersive simulations, this feedback becomes experiential to allow teachers to see and hear the consequences of their instructional moves while simultaneously accessing data visualizations that promote self-assessment and informed adjustment.

Finally, teacher analytics has become increasingly integrated into preservice and in-service teacher education programs, particularly through immersive VR-based simulations. Embedding real-time analytic insights into these simulations allows teachers to experiment, reflect, and refine their instructional strategies in authentic yet controlled settings. This integration transforms simulation from a static practice tool into a dynamic learning system where data and experience co-evolve. Such an approach not only enriches simulation-based learning but also contributes to establishing a sustainable, data-informed culture of professional development.

Despite these promising advances, current research remains predominantly centered on student data, with studies explicitly focusing on teacher analytics still relatively scarce (Sergis & Sampson, 2017; Nguyen & Karunaratne, 2024). A comprehensive framework for understanding and operationalizing teacher analytics in immersive contexts is therefore urgently needed. Key challenges include determining valid indicators of teaching competence, defining what constitutes meaningful evidence of professional growth, and ensuring that analytic insights can be ethically and effectively integrated into teacher training workflows (Hoyos & Velásquez, 2020; Saar et al., 2022). Addressing these methodological and ethical challenges will be essential for advancing learning analytics in teacher simulations from experimental prototypes to robust, scalable systems for professional learning.

*Multimodal Teacher Analytics: Integrating Behavioral and Cognitive Indicators*

Improving teacher education requires careful analysis of data generated by training programs. Such data can generally be categorized into two major types: behavioral and cognitive. The following section discusses these two dimensions in detail and examines how they can be meaningfully integrated to support comprehensive and data-informed teacher learning, as schematically illustrated in Fig. 1.

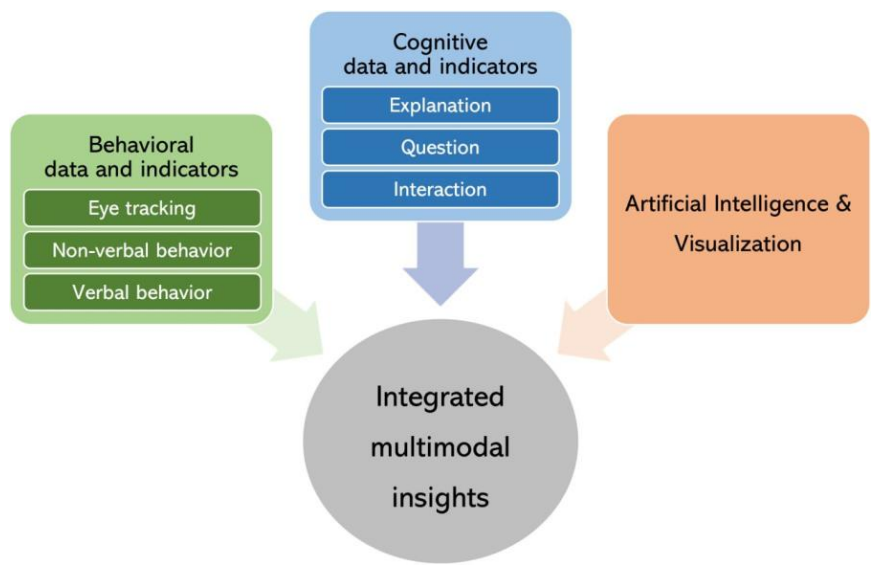

**Fig. 1** Components of multimodal teacher analytics

*Behavioral Data and Indicators*

Non-verbal behaviors are central to effective teaching and teacher–student relationships (Bambaeeroo & Shokrpour, 2017). Visual attention, posture, movement, and gesture help regulate classroom dynamics. Among these, visual perception is especially critical. Eye-tracking reveals where teachers direct their gaze and what they overlook, with studies showing experts focus more on meaningful or problematic behaviors than novices (Wolff et al., 2016; Keller et al., 2022; Kosko et al., 2024). Metrics such as fixation duration and attention dispersion indicate classroom awareness and instructional expertise. A common limitation, however, is that many studies track gaze while participants view pre-recorded videos rather than capturing real-time behavior in interactive settings.

Beyond gaze, gestures, facial expressions, and bodily movement are key non-verbal channels. Advances in computer vision and wearable sensors now enable automatic recognition and feedback on teachers' movements and body language, supporting self-reflection and improved practice (Barmaki & Hughes, 2018; Dimitriadou & Lanitis, 2025). Still, aspects of physical interaction—such as how teachers circulate and engage different student groups—remain underexplored despite their relevance for maintaining engagement and discipline.

Verbal communication is another core dimension. Speech rate, pausing, intonation, and volume affect comprehension, classroom climate, and even vocal health, and are increasingly used as evaluative features (Remacle et al., 2023; Järvinen et al., 2024; Niculescu et al., 2025). Communication style—e.g., balancing directive and facilitative discourse—also shapes student participation (Järvinen et al., 2025). Together, these verbal and non-verbal metrics depict teaching as a performative, data-traceable activity.

*Cognitive Data and Indicators*

Analyzing teachers' cognitive moves and discourse is key to judging instructional quality. Teaching comprises operations such as explaining, questioning, responding, and scaffolding (Allen & Eve, 1968); examining these reveals the instructional logic by which knowledge is organized and conveyed.

Effective explanations often use analogies to connect with learners' prior experiences or employ advance organizers (Ausubel, 1978). Accordingly, coding schemes quantify teachers' cognitive actions and discourse categories (Tong et al., 2024; Mu et al., 2025; Gjære & Mosvold,

2025). Findings commonly show that experts vary in questioning strategies, use higher-order prompts, and sustain dialogic exchanges, while novices rely more on monologic explanation and closed questions (Tong et al., 2024; Mu et al., 2025). Yet patterns can invert, and immersive simulations provide controlled contexts to examine how task constraints produce such shifts (Khoza & Magadlela, 2025).

Discourse analyses also distinguish teacher-centered classrooms—lecturing and factual recall—from student-centered settings that promote open-ended dialogue and collaborative sense-making (Mu et al., 2025). Capturing these dynamics in immersive simulations offers an analytic lens on not only content delivery but also teachers' cognitive adaptability and responsiveness to learner input.

*Multimodal Data and Indicators*

The integration of behavioral and cognitive indicators within a unified analytic framework—known as multimodal teacher analytics—offers a powerful approach for capturing the multifaceted nature of teaching. By linking data from mobile eye-trackers, audiovisual recordings, physiological sensors, and automated discourse transcriptions, teacher behaviors can be studied from complementary perspectives (Prieto et al., 2018). In addition to these hardware-intensive methods, recent semi-analytic approaches using *RayCast*-based gaze estimation within VR environments have emerged as cost-effective alternatives. By projecting virtual sightlines from the user's head or controller orientation, *RayCast* analytics approximate teachers' visual attention and enable the generation of heatmaps, fixation durations, and attention trajectories within immersive classrooms. Although less granular than optical eye-tracking, this technique provides scalable and accessible means to study attentional focus and situational awareness during teacher simulations. Such multimodal integration supports richer diagnostic and formative feedback, bridging the gap between observed actions and underlying cognitive processes.

**Empirical Cases of Data-Informed Learning through Immersive Teacher Simulations**

Today, immersive simulations instrumented with behavior, cognitive, and affective data have shifted analytics from merely documenting performance to actively supporting teacher learning–moving from description to diagnosis to real-time guidance. This section traces that role shift.

*Patterns of Teaching Strategies from Behavioral Data (Frequency and Sequence)*

How often do teachers use particular strategies, and in what order do they connect them? To address this, Marelle et al. (2025) conducted a study with preservice special education teachers using the *TeachLivE* mixed-reality classroom simulation. In the simulation, avatar students deliberately displayed disruptive behaviors, and the teachers were required to apply a range of behavior management strategies in response. The researchers documented the frequency of strategies such as providing opportunities to respond, delivering behavior-specific praise, and using token reinforcement through structured observation checklists.

Data analysis examined how the frequency of these strategies changed across simulation sessions. Teachers' use of evidence-based management practices increased significantly over time, indicating that data-informed feedback within the simulation promoted positive shifts in teacher behavior. Taken together, this case shows how immersive simulations can leverage data-driven feedback to foster meaningful behavioral growth in teacher education.

Building on these frequency-based findings, Hu et al. (2025) examined not just how often strategies appeared but how they unfolded in time. In a VR microteaching program for preservice teachers, classroom discourse was time-stamped, coded, and analyzed with lag sequential analysis (LSA) to estimate transition probabilities among moves. Compared with traditional microteaching, the VR-trained group exhibited richer and more integrated behavioral chains, including prominent sequences such as 7→2 (demonstrate through modern multimedia → ask question), 4→2 (correct and supplement → ask question), 8→1 (classroom supervise and control → explain), and 9→3 (practicing → accept and praise) (See Figure 2). These higher-prevalence paths indicate broader coverage of classroom management, reinforcement, and organizational skills, and suggest that immersive contexts cultivate productive routines (e.g., management → explanation, feedback → questioning) rather than isolated moves. Behavioral data underscore an integrated frequency–sequence perspective: what strategies teachers use matters, but when and after what they deploy them is directly tied to learning effects. If fully realized, this perspective could enable an in-simulation loop of measurement, feedback, and adjustment that makes behavioral change visible and supports deliberate practice.

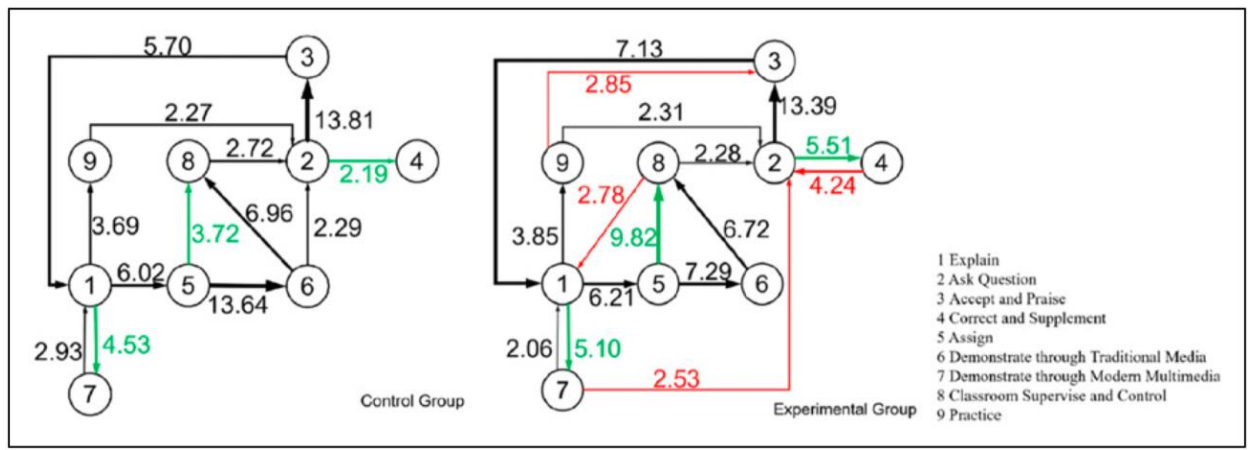

**Fig. 2** Behavioral sequence in the experimental group—significant paths from lag sequential analysis (Hu et al., 2025).

*Professional Vision from Perceptual–Cognitive Data*

Where do teachers allocate attention? To ground this question empirically, we begin with perceptual–cognitive evidence (e.g., gaze attention) that renders attention allocation observable and coachable in immersive settings. Perceptual and cognitive data, such as eye-tracking, offer insights into how teachers allocate attention and notice important classroom events. Huang et al. (2025) conducted a study with preservice teachers who engaged in a VR classroom scenario while wearing eye-tracking headsets. The system collected detailed visual attention metrics, including fixation duration, spatial dispersion, and latency in detecting salient classroom events. These data were used in real time to generate adaptive feedback, providing participants with personalized prompts about where and when to direct their attention. By presenting individualized visual attention distributions, the system enabled teachers to compare their gaze allocation against expected patterns and adjust their focus accordingly. Teachers in the adaptive feedback condition significantly outperformed those in static and no-feedback groups in both accuracy and speed of detecting student cues.

This study demonstrates how perceptual data can be translated into actionable guidance that develops teachers' professional vision, thereby supporting a skill that is otherwise difficult to teach in traditional field placements. Figure 3 illustrates this adaptive process, showing the feedback interface and corresponding heatmaps for teachers with more inert (left) versus more dynamic (right) visual movement patterns. In sum, perceptual–cognitive analytics make the invisible visible—turning patterns of visual attention into concrete, coachable skills. Yet gaze is not cognition: identical fixations may reflect different instructional intentions, depending on task and discourse context. Moving forward, the integration of gaze data with concurrent behavioral and discourse indicators within unified dashboards will allow teachers not only to see where they looked but to understand why—transforming attention metrics from descriptive observation into meaningful pedagogical reflection.

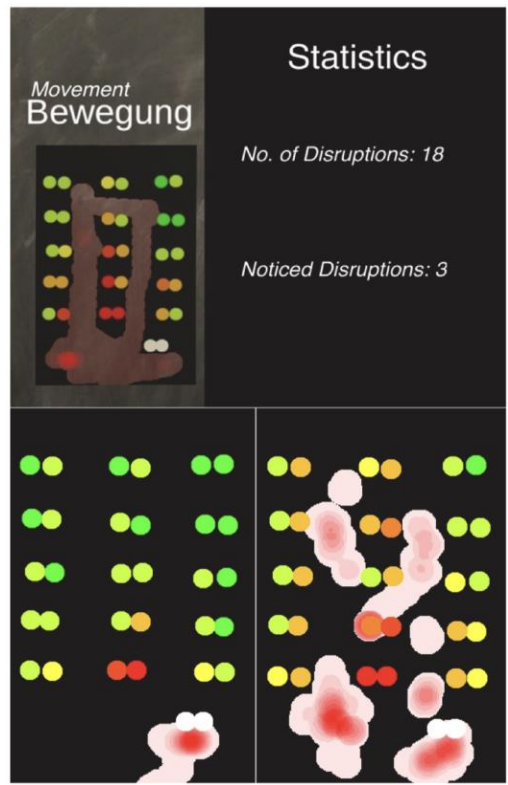

**Fig. 3** Adaptive feedback interface (top) and heatmaps contrasting inert vs. dynamic movement (bottom) (Huang et al., 2025).

*Interaction Analysis from Discourse Data*

Who do teachers interact with, and how do they interact? A growing body of work has addressed these questions across immersive and mixed-reality environments. Among representative studies, Cao et al. (2025) implemented a professional development program for in-service science teachers in Pakistan using the Mursion mixed-reality platform. Across repeated simulation sessions, avatar students presented diverse and occasionally challenging responses, requiring teachers to adjust their instructional strategies in real time. All teacher–student interactions were transcribed and systematically coded for responsiveness, with particular attention to how teachers framed discussions and adapted their instructional moves (see Fig. 4). Discourse analysis revealed that, over the course of the program, teachers increased both the frequency and sophistication of contingent responsiveness, demonstrating a stronger ability to build on student ideas. Follow-up classroom observations confirmed that these improvements transferred to authentic teaching practice, indicating that the skills developed in the simulation generalized to real-world contexts.

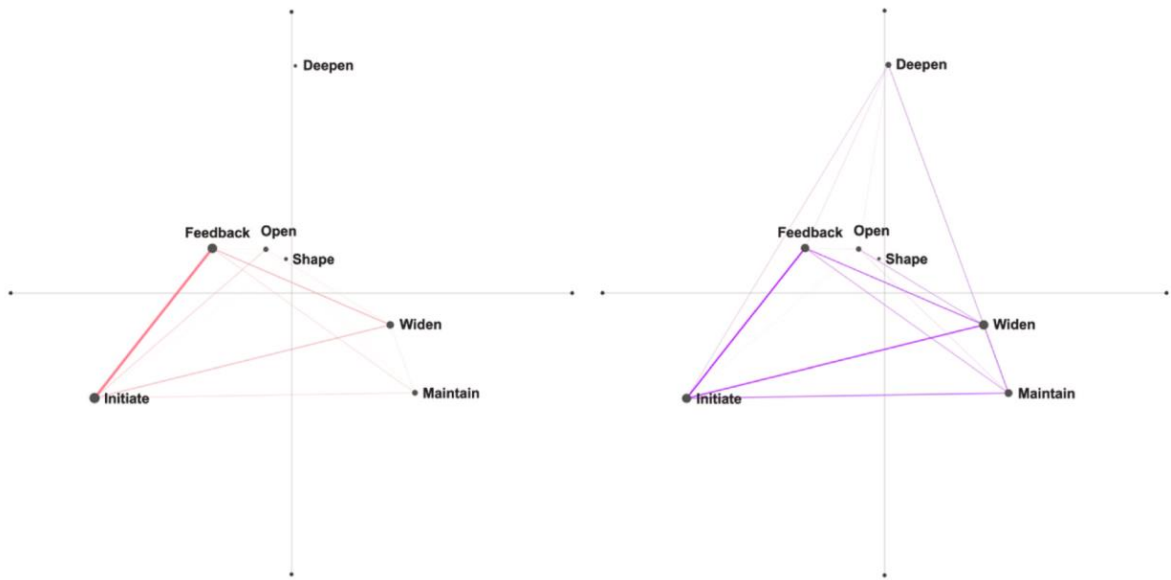

**Fig. 4** ENA (Epistemic Network Analysis) graphs (Cao et al., 2025): pre- vs. post-PD; new initiate→maintain, widen→maintain links signal more student-driven talk

This case illustrates how discourse and interaction data can serve as a powerful basis for systematic reflection and feedback. By making conversational dynamics visible, such analyses allow teachers to recognize their own patterns of turn-taking, questioning, and response framing. When integrated into interactive dashboards, these insights can move beyond post-hoc evaluation toward real-time facilitation, helping teachers visualize and refine their responsiveness during ongoing practice. In this way, discourse analytics represent a pivotal step in transforming simulation data from descriptive records into adaptive learning supports for professional growth.

Transcribing and coding teacher–student talk turns responsive teaching moves into visible indicators, establishing a cycle of measurement, feedback, and improvement. The next frontier is to extend these indicators beyond individual performance toward equity monitoring—tracking, for instance, the distribution of speaking opportunities and the balance of participation among students. However, such automation and standardization remain at an early stage. Moving forward, developing calibrated analytic thresholds, along with clear fairness and privacy guardrails, will be essential. When integrated and validated within teacher-facing dashboards, these systems can provide not only diagnostic insights but also continuous, equitable guidance for fostering inclusive classroom discourse.

*Scalable Teacher Training from Automated Performance Data*

How can analytics be embedded to make teacher training both effective and scalable? The next generation of immersive simulations increasingly integrates automated data pipelines directly into the training environment, transforming what was once a researcher-driven analytic process into real-time, teacher-facing feedback systems.

King et al. (2022) demonstrated this potential using an AI-augmented VR platform that automatically logged teacher–student interactions and applied AI-based scoring to assess procedural fidelity. Alongside observation coding and self-efficacy surveys, the study revealed significant and sustained gains in questioning-strategy fidelity and instructional confidence for teachers in AI-enhanced VR treatment. Automation substantially reduced instructor workload, enabling broader implementation across cohorts while preserving training quality.

Building on this direction, Niculescu et al. (2025) proposed an AI framework capable of extracting performance and behavioral measures from multimodal inputs—including audio, video, and gesture data—to deliver precise feedback in near real time with minimal human intervention.

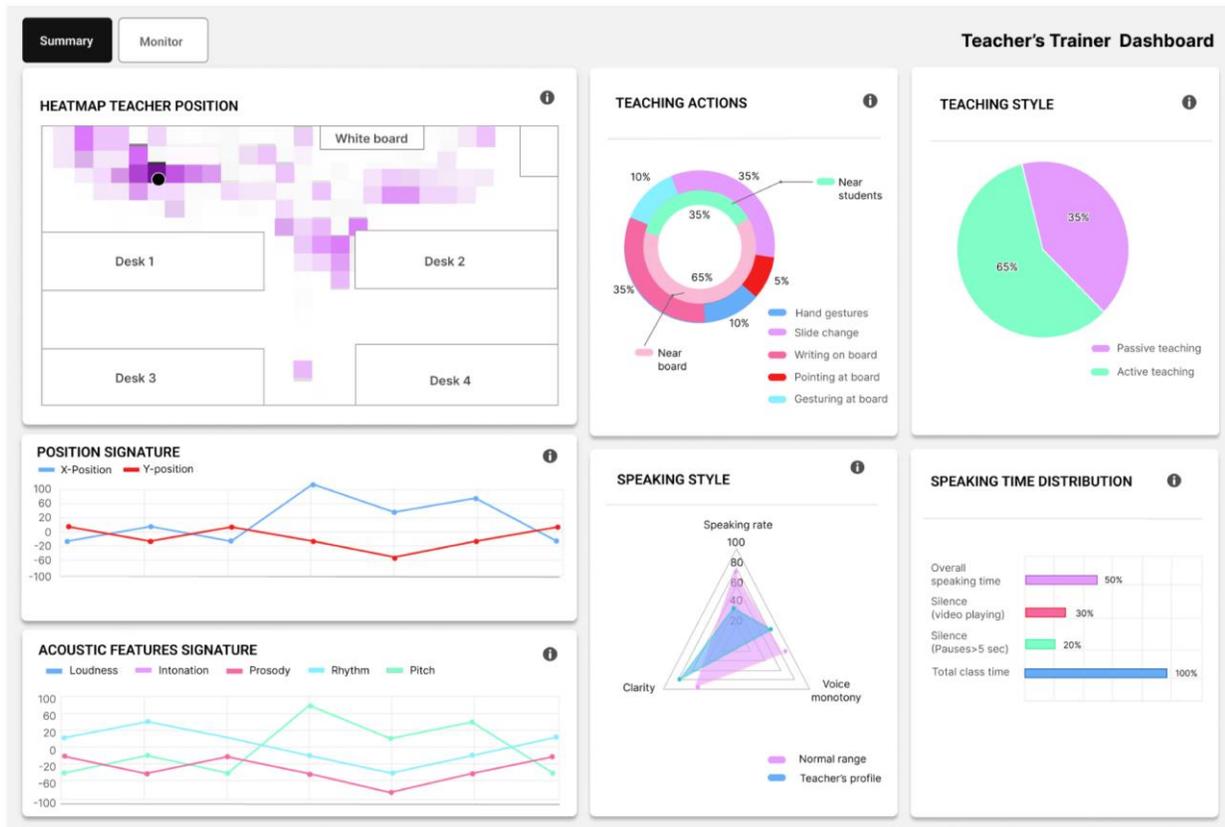

**Fig. 5** Prototype teaching analytics dashboard (Niculescu et al., 2025)

Figure 5 illustrates a prototype teacher analytics dashboard adapted from Niculescu et al. (2025). The interface consolidates multimodal indicators into a unified, non-judgmental feedback display. Key panels visualize spatial movement patterns (heatmaps), teaching actions, voice prosody, gaze or position signatures, and speaking-time distributions. Together, these components provide teachers with a comprehensive yet interpretable overview of their instructional behaviors, transforming complex analytics into actionable insights for self-reflection and improvement. Collectively, these developments signify the field's transition from post-hoc data analysis to continuous, scalable professional learning. Automated performance dashboards operationalize analytics as real-time mediators of teacher growth—turning multimodal data into accessible feedback loops that sustain deliberate practice at scale.

**Study Approach and Design**

*Research Design*

This study adopted a case study design focused on the implementation of *TeacherGen@i*, a generative AI-enhanced immersive teacher simulation (Fig. 6). We aimed to capture and analyze preservice teachers' pedagogical behaviors and discourse patterns in a controlled, yet realistic, virtual classroom environment. Our design leverages a multimodal analytics framework that integrates natural language processing (NLP) and behavioral detection to transform unstructured simulation data into actionable educational insights.

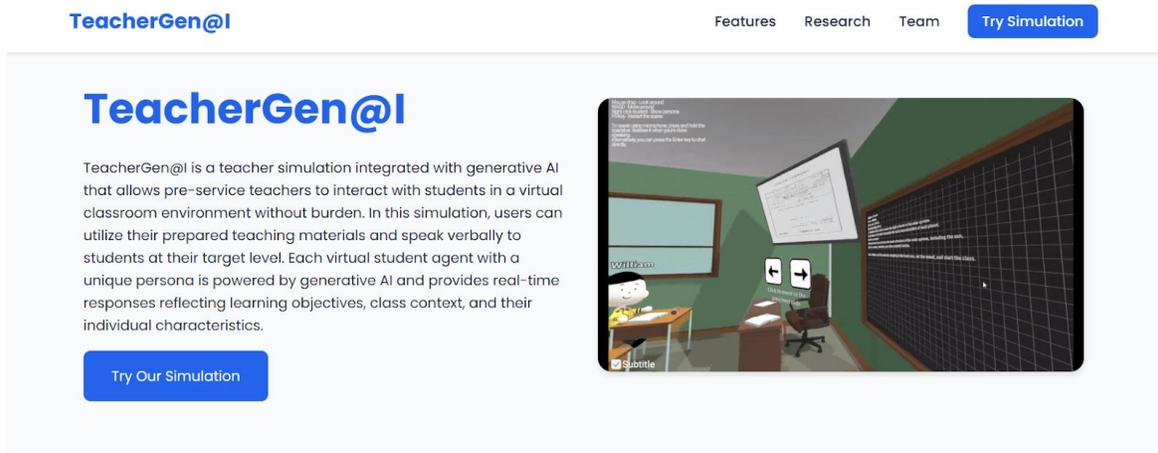

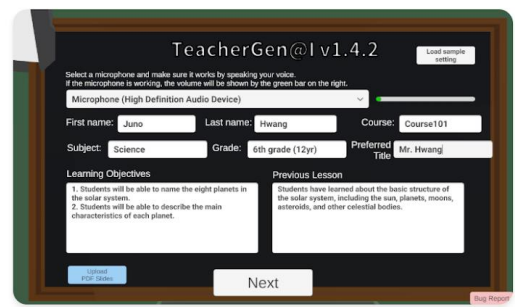

**Fig. 6** Interface of TeacherGen@I (Hwang et al., 2024)(https://teachergenai.github.io/)

*Context and Participants*

The study was conducted at University A in South Korea in the Spring 2025 semester. The context was a *Digital Education* course designed to enhance preservice teachers' digital competencies. The participants were 33 preservice teachers from the College of Education. They had varying levels of prior experience with VR technology, but had foundational knowledge in educational theory. A total of 33 simulation sessions were recorded and analyzed. To ensure ethical standards, this study was reviewed and approved by the Institutional Review Board (IRB) of The University of Alabama (Approval No. 24-02-7395). All participants provided informed consent regarding the recording of their voices, eye movements, and virtual behaviors for research purposes.

### Procedures

*Scenario Design*

Before the simulation, participants utilized the TeacherGen@i system to design specific teaching scenarios. They set learning objectives, adjusted class level (i.e., students' concentration level), and structured classroom situations aligned with the national curriculum standards.

*Immersive Simulation*

Participants completed the simulation within a web-based VR environment. They conducted a teaching session where they interacted with AI-driven student agents. These agents were powered by a large language model (LLM) to generate adaptive responses and unique personas tailored to the participants' designed scenarios.

*Analysis*

After the simulation, participants reviewed their lesson through a prototype dashboard. However, the primary focus of this chapter is on analyzing the data generated during the simulation phase.

### Data Collection

Data were collected in real-time through the VR system and consisted of two main streams. The first was *verbal data*, where all verbal interactions were captured using a Whisper-based Speech-to-Text (STT) system to ensure accurate transcription of real-time verbal interactions. The second was *multimodal behavioral data*; the system logged timestamped events of teacher and student actions, while eye-tracking sensors and gaze direction at each frame.

*Data Analyses*

The collected data underwent a multi-stage analysis. For *discourse analysis*, transcribed texts were classified using a BERT-based model according to Bloom's cognitive taxonomy, followed by dimensionality reduction using t-SNE to visualize these embeddings in a semantic space. To capture *behavioral and perceptual dynamics*, Lag Sequential Analysis (LSA) was conducted to identify significant interaction flows (e.g., Teacher → Student). At the same time, gaze analysis processed eye-tracking data to generate spatial heatmaps and examine attentional dispersion.

**Findings**

In this section, we present the results extracted from the multimodal data—discourse, behavior, and gaze—collected from the TeacherGen@i simulation. The findings are organized to illustrate the cognitive distribution of interactions, temporal behavioral patterns, and visual attention metrics.

*Cognitive Distribution of Pedagogical Discourse*

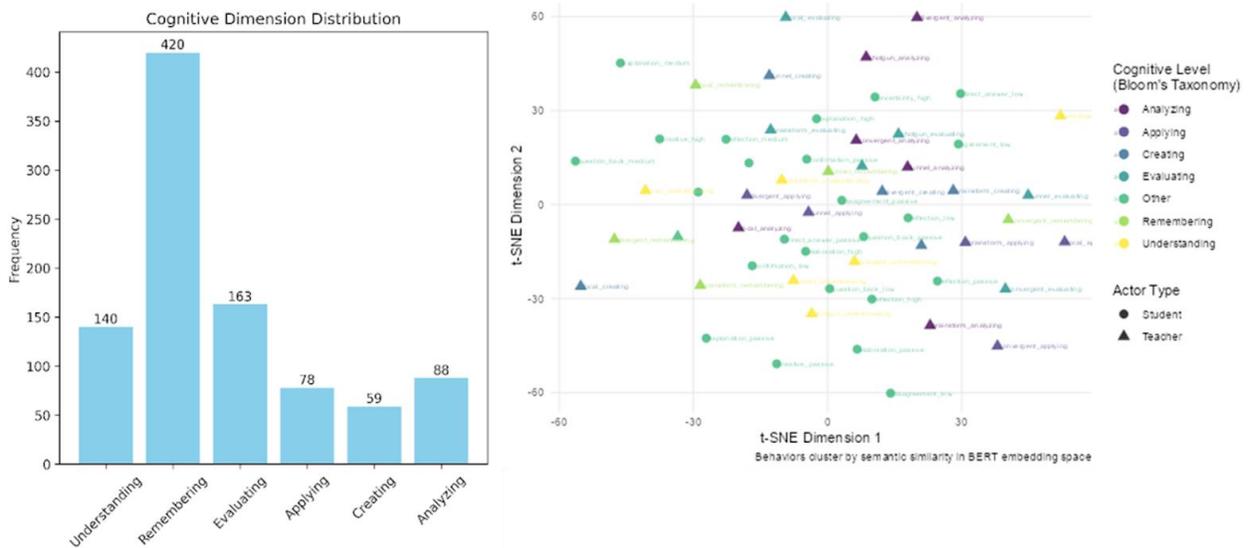

**Fig. 7** Cognitive Dimension Distribution and BERT Embedding t-SNE Semantic Space of Verbal Data in Immersive Teacher Simulation

A total of 33 dialogue sessions involving preservice teachers were analyzed, comprising 1,269 teacher–student interaction turns with an average conversation length of 38.5 turns per session. The left panel of Fig. 7 shows the frequency distribution of cognitive dimensions derived from BERT-based classification of preservice teachers' verbal data. It illustrates a predominance of Remembering and Understanding questions during instructional interactions. This pattern shows that preservice teachers frequently employed recall-oriented discourse patterns, primarily focusing on confirming facts or checking basic comprehension during instructional exchanges. The right panel of Fig. 7 visualizes the same data in a BERT-based behavioral semantic space using t-SNE dimensionality reduction. Each point represents a coded behavior, with color denoting its cognitive level (according to Bloom's taxonomy) and shape representing actor type (student or teacher). Behaviors cluster according to semantic similarity, revealing distinct meaning zones where higher-order cognitive processes (Analyzing, Evaluating, Creating) diverge from lower-order processes (Remembering, Understanding).  To sum up, the distribution highlights the overall cognitive tendencies of preservice teachers, while the semantic space exposes latent relationships and overlaps between teacher and student utterances. This combination demonstrates how linguistic embeddings can bridge quantitative classification and qualitative interpretation, mapping the semantic topology of teaching behaviors within immersive, AI-mediated learning environments.

*Sequential Dynamics of Interaction*

To further examine the temporal dynamics of instructional dialogue, Sequential Analysis and Lag Sequential Analysis (LSA) were conducted. These analyses identified 215 significant behavioral patterns, revealing structured interactional flows across teacher–student exchanges: (1) Teacher-to-Teacher (T→T) interactions exhibited cognitive scaffolding chains (69 patterns; 34.8%), such as *t_focal_understanding* → *t_funnel_analyzing*. (2) Student-to-Student (S→S) sequences (110 patterns; 55.6%) reflected critical thinking progression, including *s_disagreement_low* → *s_creative_high*. (3) Student-to-Teacher (S→T) feedback loops (16 – patterns; 8.1%) captured question-triggering processes, such as *s_creative_high* → *t_divergent_evaluating*. These sequential mappings illuminate how generative AI–driven student agents can elicit increasingly complex cognitive and dialogic patterns from preservice teachers.

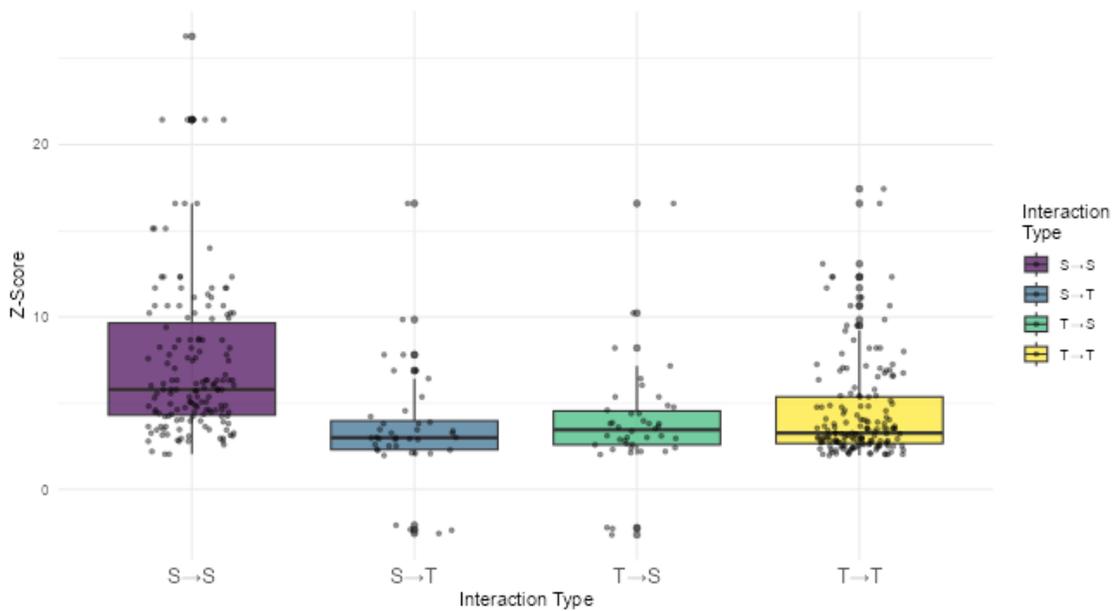

**Fig. 8** Lag Sequential Analysis: Z-Score Distribution by Interaction Type.

Figure 8 shows the distribution of Z-scores across four interaction types (S→S, S→T, T→S, T→T). Student-to-student interactions display the highest median Z-scores and variance, indicating dense, self-sustaining discourse chains where elaboration and co-reasoning frequently co-occur. Teacher-mediated interactions (T→S and T→T) exhibit moderate but stable patterns, suggesting structured scaffolding sequences that regulate peer discourse. Student-initiated turns (S→T) appear less frequently but produce notable spikes in Z-scores, marking moments of cognitive disequilibrium that often trigger teacher reflection or adaptive feedback. Altogether, these distributions indicate that immersive simulations encourage a hybrid interaction structure—both teacher-guided and peer-driven—where dialogic reasoning becomes measurable.

Furthermore, Figure 9 provides a fuzzy network visualization of the 49 strongest sequential relationships ($Z > 10$). Nodes represent coded behaviors—teachers (red) and students (blue)—while directed edges encode temporal dependencies, with line thickness corresponding to Z-score magnitude and color signifying interaction type (S–S, S–T, T–S, T–T). This fuzzy network does not impose a linear hierarchy of cause and effect but instead reveals the probabilistic and overlapping nature of instructional sequences. Multiple links may emanate from a single node, showing that the same teacher move (e.g., funnel_analyzing) can lead to different student responses depending on context and preceding actions. Such visualization captures the inherent variability and adaptivity of teaching processes—phenomena that traditional linear statistics tend to flatten. By making this "fuzziness" visible, the network serves as a diagnostic and reflective tool, helping researchers and teacher educators recognize how instructional strategies branch, converge, and evolve within immersive classrooms.

Together, Figures 8 and 9 demonstrate how sequential analytics and fuzzy networks jointly illuminate the temporal logic of instruction—revealing not only the frequency of actions but also their dynamic interdependencies. In this sense, TeacherGen@i transforms teaching from a black-box performance into a traceable, analyzable, and ultimately improvable process of data-informed pedagogical decision-making.

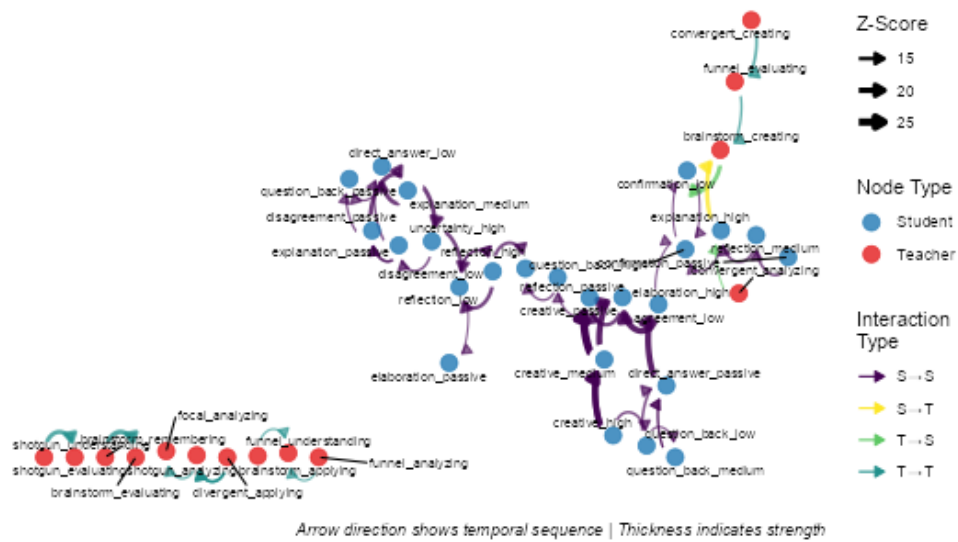

**Fig. 9** Network of Strongest Sequential Relationships in Immersive Teacher Simulation (TeacherGen@i)

*Multimodal Visualization: Gaze and Dashboard*

To explore how multimodal analytics fosters reflective teaching, we developed a prototype dashboard visualizing data from TeacherGen@i, as shown in Fig. 10. Designed for formative analysis rather than large-scale training, this interface serves as a tool for examining virtual teaching behaviors and evaluating their impact on professional reflection.

The prototype integrates several data streams—verbal interaction logs, coded teaching actions, sequential behavior patterns, and gaze-tracking information—to generate post-session visual feedback on teaching presence and classroom awareness. Collectively, these multimodal indicators provide a composite picture of how teachers attend, act, and interact in real time, demonstrating the potential of analytics to move beyond post-hoc diagnosis toward actionable feedback and self-regulated improvement.

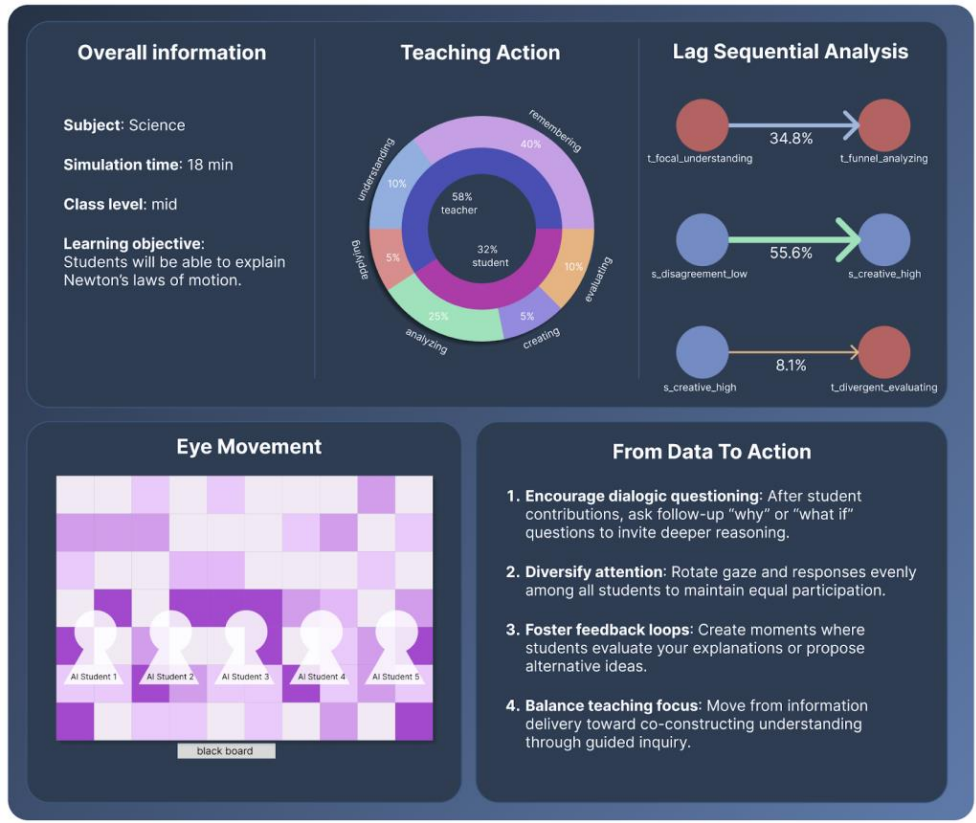

**Fig. 10** Dashboard Prototype

Within the dashboard, contextual information about each simulation session, such as subject area, duration, learning objectives, and class level, provides interpretive scaffolds for understanding behavioral and cognitive patterns. Visual analytics further display the distribution of teacher and student behaviors categorized by cognitive process levels based on transcribed dialogue logs coded using a modified Bloom's taxonomy. Complementing these representations, LSA visualizations reveal statistically significant behavioral transitions, including teacher-to-teacher scaffolding, student-to-student reasoning, and student-to-teacher feedback loops. These sequences, derived from timestamped interaction logs, are represented as directional links with line thickness indicating transition frequency, thereby highlighting the dynamic structure of classroom discourse and response patterns.

In parallel, eye movement visualization employs gaze data captured through the VR headset's integrated eye-tracking sensors, which record head position and gaze direction at each frame. By projecting these vectors onto the virtual classroom space, the system identifies which virtual student each teacher attends to and for how long, generating spatial heatmaps that depict attention allocation patterns across individual students. Alongside these spatial visualizations, saccade analysis offers a parallel analytic approach that quantifies attentional dispersion by revealing whether teachers maintain focused engagement or exhibit fragmented scanning behaviors. Together, these gaze-based visualizations allow users to examine how visual attention aligns—or fails to align—with key instructional moments, offering a powerful cue for improving classroom awareness. The feedback panel then synthesizes these multimodal insights into concise, teacher-friendly guidance. By providing individualized reflections on attention balance, interactional equity, and feedback strategies, the dashboard bridges the gap between complex analytic outputs and pedagogically meaningful interpretation.

An additional design priority is ensuring accessibility and inclusivity. Immersive analytics systems should be adaptable for teachers with varying technological proficiency, language backgrounds, and physical abilities. Low-cost headsets, multilingual interfaces, and adaptable input modes (e.g., controller-free or voice-based interaction) can broaden participation across diverse teacher populations. Furthermore, analytics dashboards can integrate equity indicators, such as speaking-time ratios or gaze balance, to support inclusive classroom practices and raise teachers' awareness of implicit bias.

Taken together, the VR Simulation Dashboard represents a conceptual step toward embedding multimodal learning analytics directly into immersive teacher education. By integrating behavioral, discourse, and perceptual data within a unified visualization framework, the prototype demonstrates how immersive analytics can evolve from fine-grained, researcher-centered observations to real-time, teacher-facing reflection systems. This progression underscores the transformative potential of generative AI and multimodal analytics in advancing scalable, data-informed teacher professional development.

**Understanding Innovative Practices in Teacher Education**

The multimodal analyses conducted in this study—spanning cognitive discourse, sequential interaction patterns, and visual attention—provide insights that extend beyond performance description. By rendering invisible instructional moves visible, TeacherGen@i transforms teaching from a black-box phenomenon into a traceable, analyzable, and improvable process. This visualization enables preservice teachers to objectively examine their questioning strategies and interaction flows, thereby fostering analytics literacy and supporting evidence-based pedagogical decision-making. Furthermore, these findings suggest a curricular shift toward iterative "simulation–analysis–redesign" cycles, where data dashboards serve as essential artifacts for cultivating professional vision and developing instructional routines within teacher education.

*Interpreting Cognitive and Behavioral Findings*

The analysis of the Cognitive Distribution of Pedagogical Discourse (the left panel of Fig. 7) revealed a predominance of Remembering and Understanding questions. This finding aligns with prior research suggesting that preservice teachers frequently adopt convergent questioning styles in early instructional experiences (Mu et al., 2025; Tong et al., 2024). This suggests that immersive simulations, although complex and dynamic, do not in themselves guarantee higher-order pedagogical reasoning unless deliberate scaffolds are in place. The t-SNE semantic space further reveals densely clustered lower-order moves, while higher-order behaviors appear as dispersed, low-density points, a structure typical of novice instructional cognition.

Beyond this categorical distribution, the integration of linguistic embeddings (the right panel of Fig. 7) highlights deeper interpretive affordances. These approaches illustrate how BERT captures both categorical and contextual features of pedagogical discourse. This combination demonstrates how linguistic embeddings can bridge quantitative classification and qualitative interpretation, mapping the semantic topology of teaching behaviors within immersive, AI-mediated learning environments.
These representational qualities reinforce the value of embedding-based discourse analysis in teacher education research, offering ways to surface latent structures in instructional communication.

The sequential analyses presented in Fig. 8 complement the cognitive findings. Structured T→T scaffolding chains and robust S→S reasoning sequences indicate that simulations can evoke patterned dialogic routines even among novices. Meanwhile, the intermittent but high-impact S→T initiation turns resemble what noticing and adaptivity literature describe as cognitive disequilibrium, moments that invite teachers to shift questioning strategies or instructional focus (Keller et al., 2022; Wolff et al., 2016).

Lastly, the fuzzy sequential network (Fig. 9) illustrates how instructional behavior unfolds along branching pathways rather than linear cause–and–effect sequences. This pattern resonates with literature on adaptive expertise, demonstrating that preservice teachers do not simply follow prescriptive routines but generate variable responses shaped by local context and generative student agent input. Finally, gaze dispersion and heatmap patterns reveal novice tendencies toward selective noticing, often prioritizing visually salient or centrally positioned student avatars. This echoes established findings in professional vision research (Kosko et al., 2024), which document uneven attention allocation as a developmental marker in early teaching practice.

*Connections to Research on Teacher Learning*

The findings extend existing work on immersive teacher simulations by showing how professional vision, instructional adaptivity, and cognitive scaffolding can be examined through multimodal analytic lenses (Huang et al., 2025; Cao et al., 2025). While prior research demonstrates the realism and authenticity of VR classrooms, this study deepens the analytic layer by integrating discourse patterns, gaze signatures, and sequential structures.

Moreover, the combination of categorical coding and embedding-based semantic modeling demonstrates the potential of generative AI methods to reveal underlying meaning structures in teacher–student discourse. By bridging quantitative classification and qualitative interpretive possibilities, these linguistic embeddings position immersive simulations as fertile environments for investigating the semantic organization of instructional behavior.

The dashboard prototype (Fig. 10) contributes to ongoing discussions around analytics-informed teacher inquiry (Saar et al., 2022). By structuring multimodal data into visual representations, behavior graphs, attention heatmaps, and sequential interaction flows, it illustrates how simulations can evolve beyond performance rehearsal toward sustained, evidence-based reflection.

**Key Takeaways of the Innovation Journey**

This chapter traced how immersive teacher simulations have evolved from static practice tools into dynamic ecosystems that integrate multimodal analytics and adaptive feedback. Through empirical cases and design-based inquiry, we showed how behavioral, cognitive, and affective data combine to enrich data-informed professional learning.

The next generation of immersive learning platforms will be characterized by their ability to offer automated, intelligent, and personalized feedback in real-time. A critical challenge for teacher education is bridging the gap between practice in a simulated environment and performance in a real classroom (Jamil & Isiaq, 2019; Kaufman & Ireland, 2016). The promise of analytics-enhanced simulations lies in making data actionable and creating a clear pathway from insight to implementation (Ley et al., 2023; Moon et al., 2024). Post-simulation dashboards will offer more than summary scores; they will provide meaningful visual evidence of pedagogical patterns. For example, teachers may see a heat map of their gaze revealing an unconscious tendency to focus on one side of the classroom, or a breakdown showing that they ask predominantly closed-ended questions.

This kind of data-informed self-reflection catalyzes meaningful change in instructional design (Connolly & Ceallaigh, 2023). Grounded in established instructional design models, teachers can use specific and objective insights to revise their lesson plans. This establishes a powerful, evidence-based cycle of professional development: practice within the simulation, analyze objective performance data, reflect on pedagogical habits, redesign the lesson plan with targeted strategies, and implement it in a real classroom.

Within this cycle, the integration of gaze and discourse visualizations also supports equity-oriented reflection. Low-cost headsets, multilingual interfaces, and adaptable input modes can expand access to immersive professional learning environments. Meanwhile, equity indicators such as speaking-time ratios and gaze balance help teachers identify potential implicit biases. These insights encourage more balanced questioning strategies and more equitable classroom engagement patterns across diverse student groups. In conclusion, TeacherGen@i demonstrates how learning analytics can go beyond mere assessment and become a foundational element of ongoing reflective practice and professional growth.